Paper title

Mechanical and Optical Characterization of a Suspended Core Fiber Exhibiting Fundamental-Mode Cutoff Wavelength in Presence of Nanoscale Air Holes in the Core Region.

Authors:

    Dihan Hasan

    M. Shah Alam

Authors Affiliation:

    School of Electrical Engineering and Computer Science, Oregon State University, Corvallis, OR, 97331, US.
    hasandihan@gmail.com

    Professor, Department of Electrical and Electronic Engineering, Bangladesh University of Engineering and Technology (BUET), Dhaka-1000, Bangladesh
    shalam@eee.buet.ac.bd

Corresponding Author:

    Dihan Hasan
    School of Electrical Engineering and Computer Science, Oregon State University, Corvallis, OR, 97331, US.
    Email: hasandihan@gmail.com

**Abstract:** In this work, mechanical and optical characterization of a new type of suspended core fiber (SCF) has been performed. The proposed SCF along with additional central air holes exhibits an unusual property of fundamental mode cutoff at short wavelengths. Two variants (single hole and double hole) of design under two different fiber platforms ($SiO_2$ and $As_2Se_3$) have been considered to develop better insight into the mechanical and optical properties of the structure deploying plane strain and full vector eigen analysis, respectively. Dependence of thermal stress likely to be present in such nanostructured fibers on fiber materials and geometries are thoroughly studied. A relation between the cutoff condition and geometrical parameters of the design has been obtained with reference to characteristic decay length which nullifies the possibility of any numerical artifact. In addition, improvement of fiber birefringence and evanescence sensing capability due to the presence of such air holes in the guided region has been theoretically demonstrated. Effects of such air holes on fiber properties like GVD and mode mismatch are also studied with a relation to power confinement inside such nano scale regions.

## 1. Introduction

Suspended core fibers (SCF) have drawn the attraction of the researchers very recently due to its unique potential of maintaining significantly higher light confinement in sub-wavelength guided region compared to numerous variants of complex shaped photonic crystal fibers (PCFs) proposed so far. High numerical aperture and small mode field diameter along with low confinement loss of SCFs have made these structures a promising solution for realizing all-fiber nonlinear devices [1]. However, light confinement in nanoscale core waveguide is fundamentally limited by diffraction which spreads light from the guided region, and therefore, increases the waveguide losses significantly. At the same time, it has been demonstrated recently that, optical energy can be trapped within a sub-wavelength air hole inside the core of a PCF resulting remarkably lower attenuation [2]. Besides, enhancement of light intensity within a nano sized air hole extruded through a high index nanowire has been reported very recently in [3]. Although, fabrication of such nanoholes within the core is still emerging, engineering merits of such region on fiber group velocity dispersion have already been numerically studied extensively in [4], [5].

In this work, we have qualitatively investigated the effects of incorporating such nanohole inside the core of an $As_2Se_3$ SCF on its mechanical and optical properties. Significant field enhancement being governed by the ratio of $\varepsilon_{SiO_2}/\varepsilon_{Air} \approx 2$ in a hexagonal PCF with a minimum nanohole diameter of 110 nm has been experimentally confirmed in [6]. In this work, we have dealt with two different geometries, one with a single nanohole (Single Hole SCF (SH-SCF)) and the other with two nanoholes (Double Hole SCF (DH-SCF)) on two completely different fiber material platforms (Chalcogenide and Silica) for reasoning the underlying physics with sufficient comparisons over a broad range of operating wavelengths (0.6 µm -2.5 µm). In fact, such concept of introducing air hole region within the core has been greatly actuated by the recent successes to confine light in the low index region of different variants of single and multiple slot waveguides [7]-[9]. In this work, we are proposing such idea for SCF as a promising route to control modes of propagation, tune birefringence and improve evanescence sensing capability of SCF.

In section 2, the design technique of the proposed structure has been explained with all necessary geometrical parameters. Possible fabrication methods to realize the proposed structure have also been studied to justify the theoretical investigation. In section 3, the methods of analysis for stress calculation and optical characterization have been discussed in brief. Finally, in section 4, we have performed a detailed analysis of the simulation results.

**2. Design of the structure**

In Fig. 1(a) and 1(b), the two proposed fiber structures have been shown. The maximum effective core diameter of our structure has been set to 1.35 µm unless otherwise specified. The effective core diameter can be defined as the maximum diameter of the circle that can fit within the structured core region. Structure with minimum core diameter of 920 nm in silica background has been reported to be fabricated already [6]. The design in Fig. 1(b) can be considered analogous to a multiple slot waveguide where the parameter $d_{sep}$ has been set to 400 nm at a certain tapered condition. The two designs differ from each other in term of circular symmetry of the core region. One crucial challenge for the design is to develop a mathematical model for forming the deformed air holes surrounding the core region. In previous works, 3 or 4 large air holes have been proposed for SCF where a Y-shaped fundamental mode profile is maintained. However, to ensure circular fundamental mode profile for reducing the mode mismatch effect, we have considered 8 deformed air holes surrounding the core [10]. In this design, each hole is formed by the combination of a circular hole and a second degree Bézier curve [11]. Different geometrical condition for the core can be obtained by manipulating the control points of Bézier curve [26]. A quadratic Bézier curve is the path traced by the function B($t$) as given by,

$$B(t) = (1-t)^2 P_0 + 2(1-t)t P_1 + t^2 P_2, \ t \in [0,1], \qquad (1)$$

where the given points are $P_0$, $P_1$, and $P_2$ as shown in Fig. 1(c). The curve departs from $P_0$ in the direction of $P_1$, then bends to arrive at $P_2$ in the direction from $P_1$. In other words, the tangents at $P_0$ and $P_2$ both pass through $P_1$. In Fig. 1(d), we have shown the geometrical formation of each air hole. The Bézier curves $P_1Q_2P_3$ and and $P_1Q_2'P_3$ have been shown in Fig. 1(d) with control points $P_2$ and $P_2'$. These two curves denote two different suspended conditions of the core. Here, we have defined the suspension factor (SF) of the core as $OP_2/OC$. In Fig. 1(d), C is the center of the geometry, and O is the center of the circular hole. In this proposed design, allowable SF is spanning from 0.8 to 1.7. The Air holes become completely circular for SF=0.8. Maximum allowable SF is limited by the struts width ($w$) in between two air holes. which can also be treated as fabrication parameter. Here SF controls the effective core diameter of the SCF as well. With the increase of SF of the core diameter decreases and vice-versa. We have observed the effects of varying SF from 1.3 to 1.7 in this work.

Various fabrication techniques including extrusion, stacking, drilling, casting are adopted for suspended core fiber fabrication [1], [12]-[14]. However, presence of central air hole may pose difficulties to realize proposed fiber structures. By customizing the complex preform extrusion technique as reported in [15], the proposed structure may be fabricated with less air hole collapse ratio. Lower melting point of $As_2Se_3$ should also provide advantage over Silica if this technique is adopted. However, maintenance of thin and long strut is a challenging issue that needs to be addressed while fabricating a SCF with nanoscale core size. Successful implementation of strut width of several tens of nanometers has been reported in [16]. In this theoretical analysis, minimum strut width (w) considered is around 45 nm. Besides, successful fabrication of air hole of 20 nm diameter within an $As_2S_3$ nanowire has also been confirmed in [3] very recently. Hence, it can be expected that, the proposed structure may be realized by utilizing the latest fiber fabrication technologies.

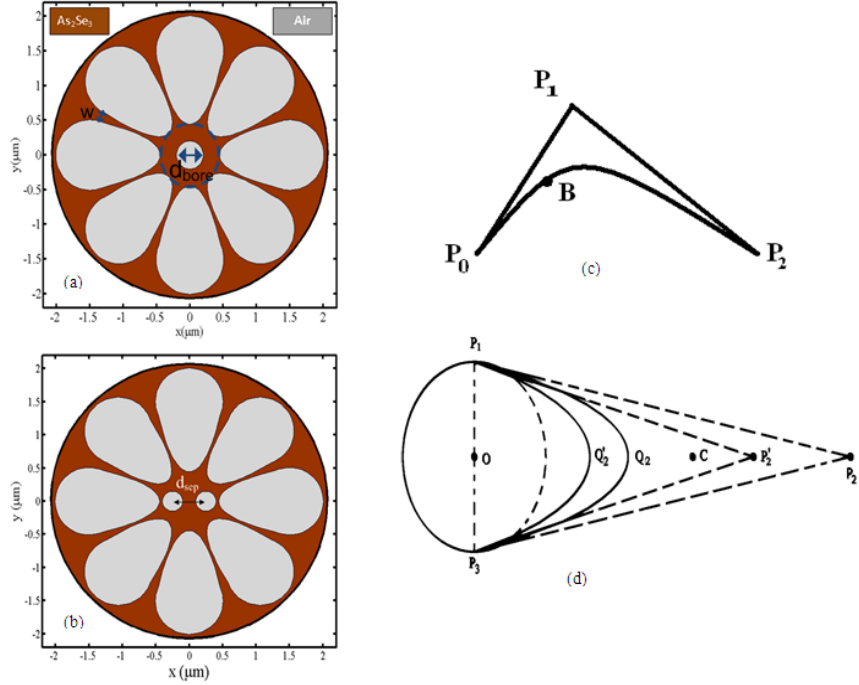

Fig.1. (a) Single Hole SCF (SH-SCF) (b) Double Hole SCF (DH-SCF). The dotted circle illustrates the effective core size. $d_{bore}$ is the centre hole diameter for both the designs. $d_{sep}$ is the centre hole separation in DH-SCF. $w$ is the strut width of the SCFs. (c) Quadratic Bézier curve (d) Formation of air holes with Bézier curves.

### 3. Method of analysis

The 2D Finite Element Method (FEM) has been employed to perform the numerical analysis considering triangular elements with adaptive meshing throughout the work [17].

#### 3.1 Plane strain analysis

The stresses resulting from thermal expansion and/or external forces in any structure can be calculated according to the following equation [23],

$$-\nabla\sigma = -\nabla\left(\begin{bmatrix}\sigma_x\\\sigma_y\\\gamma_{xy}\end{bmatrix} - \begin{bmatrix}\alpha\\\alpha\\0\end{bmatrix}(T-T_{ref})(1+v)\right) = F \text{ where } \sigma = \textbf{De} \qquad (2)$$

Here, $\sigma$ is the stress tensor, $\varepsilon_x, \varepsilon_y$ are the normal strain components and $\gamma_{xy}$ is the shear strain component. $\textbf{D}$ is the elasticity matrix describing an isotropic material using Young's modulus $E$ and Poisson's ratio $v$. $\alpha$ is the thermal expansion coefficient, $F$ is the force, $T$ is the tapering temperature, and $T_{ref}$ is the room temperature (25°C). Longitudinal strain component is neglected in our analysis. After calculating the stress components, we have studied the Von Mises stress $\sigma_v$, a quantity to estimate the yield criteria, a measure of the threshold for structural failure. In case of plane stress, it can be defined as [23],

$$\sigma_v = \sqrt{\frac{\sigma_x^2 + \sigma_y^2 + (\sigma_x - \sigma_y)^2}{2}} \qquad (3)$$

Here, $\sigma_x$ and $\sigma_y$ are the normal stress components. We have considered two different fiber materials, $As_2Se_3$ and $SiO_2$ for understanding the effects of fiber materials on structural properties. Such comparative analysis will a lay a ground to trade-off between mechanical strength of fiber and limitations of fabrication. In table I, mechanical and stress optical parameters for both the materials are listed for a comparison [18]. The material parameters are assumed to be constant at the room temperature.

TABLE I
STRESS OPTICAL DATA FOR $As_2Se_3$ and $SiO_2$

| Parameters | $As_2Se_3$ | $SiO_2$ | Air |
|---|---|---|---|
| Young's modulus $E$ [Pa] | 18×10⁹ | 78×10⁹ | 1.42×10⁵ |
| Poisson's ratio $v$ | 0.266 | 0.17 | 0.186 |
| Melting point (°C) | 180 | 1100 | - |
| Thermal expansion coefficient $\alpha$ [K⁻¹] | 15×10⁻⁶ | 0.55×10⁻⁶ | 3.43×10⁻³ |

*3.2 Optical Mode analysis*

A full vector FEM has been adopted to obtain the modal characteristics of the SCFs. The effective index is obtained from the eigen value equation corresponding to the Helmholtz equation written in the form,

$$\nabla \times (n^{-2} \nabla \times \mathbf{H}) - k_0^2 \mathbf{H} = 0. \tag{4}$$

In this work, fundamental $HE_{11}$ mode has been considered unless otherwise specified. Respective Sellmeier coefficients have been incorporated at different wavelengths during the analysis [19]. Besides, to obtain the overlap integral for estimating mode mismatch loss ignoring any fresnel reflection at the interface of the fiber and air, we have employed the following equation [20],

$$\eta = \frac{|\int E_1^* E_2 dA|}{\sqrt{\int |E_1|^2 dA \int |E_2|^2 dA}} \times 100. \tag{5}$$

Here, $E_1$ and $E_2$ are the complex electric fields of a Gaussian optical beam and waveguide fundamental mode, respectively, and the integration is to be taken over the whole cross section. The beam profile has been defined considering an x-polarized TE excitation as,

$$E_1 \approx E_x e^{-\frac{r^2}{2a^2}} \tag{6}$$

where $a$ is the Gaussian modal radius of the beam and $E_x$ is the incoming modal amplitude. The calculated mode mismatch loss is an estimation of input coupling efficiency here. Finally, to evaluate the sensitivity of the fiber, we have considered the following sensitivity equation [28],

$$\%r = \frac{n_r}{n_e} \frac{P_{sample}}{P_{total}} \times 100, \tag{7}$$

where $P_{sample}$ is the total power confined in the sample material and $P_{total}$ is that in the whole cross section and $n_r$ and $n_e$ are the refractive indices of the sample material and the effective index of the corresponding mode, respectively. During such analysis, full vector nature of propagating fields has not been considered assuming that the $E_z$ component of the guided mode will not affect the qualitative conclusions reached.

## 4. Results and discussion

### 4.1 Thermal stress

In Fig. 2(a) and 2(b), Von Mises stress distributions for both the structures have been shown. We have considered two practical outer clad diameters (10 μm and 20 μm) for including the effect of cladding layer thickness on mechanical properties of such nanoscale core fiber [21]. One can observe that, the suspended holes are subject to higher stress than any other region due to small strut width of the structures and high thermal expansion coefficient mismatch between air and fiber materials. Effects of including air holes within the core are illustrated in Fig. 2(a) and Fig. 2(b). Variation of peak Von Mises stress, which is a function of fiber material properties and drawing temperature, along the x axis can be attributed to large thermal expansion coefficient mismatch between $As_2Se_3$ and silica. However, effects of such mismatch have been greatly compensated by the large contrast between the drawing temperatures for the two chosen materials. In Fig.3 (a), we have shown the variation of the magnitude of maximum stress generated in the structures with suspension factor under different conditions. Thermal stress increases with the increase in SF as the deformed air holes are arranged more compactly while the average stut width is reduced. It is to be noted here that, mode confinement in both types of SCF can be greatly enhanced by increasing SF. Also, the maximum stress in DH-SCF structure is slightly larger than in SH-SCF structure over the range of SF considered here. This is due to the presence of two holes in DH-SCF which arise additional stress components. In our study, outer clad diameter and central hole ellipticity (e=0.5) do not have significant impacts on the maximum stress generated. Although the fabrication of small core SCF has been reported already, we have numerically compared the consequences of fiber materials and fiber structure on fiber mechanical strength for the first time [22]. In addition, the proposed central hole regions which are useful to control the mode confinement, evanescence sensing and birefringence have been found to be under high thermal stress as shown in Fig. 3(b). Average hole stress calculated by integrating Von Mises stress present at the interface of air hole and fiber material decreases with the increase in suspension factor in both types of structures as shown in Fig. 3(b). The distance between a centre hole and the cladding air holes decreases with suspension factor which eventually reduces the magnitude of stress generated at the interface of a centre hole and the core region. It is to be noted also in Fig. 3(a) and 3(b) that, maximum stress is generated at the strut regions for any value of SF. However, variation of thermal expansion mismatch between the air and fiber material has been found to affect the average hole stress significantly. In our study, central holes are more prone to thermal stress when the fiber material is $As_2Se_3$ which requires significantly lower tapering temperature.

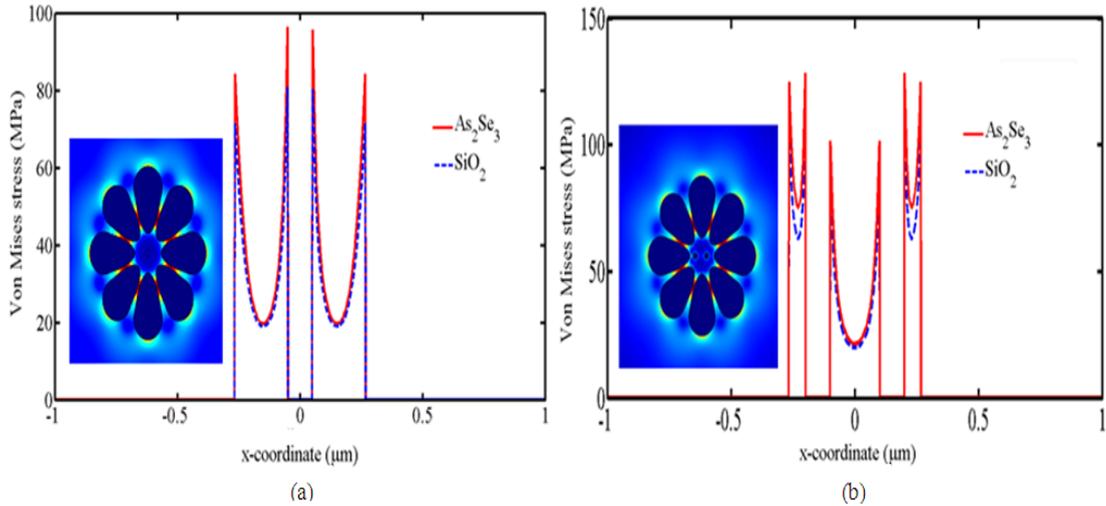

Fig. 2. Effect of nanoholes on Von Mises stress in (a) SH-SCF (b) DH-SCF, for two different materials at $d_{bore}$ =100 nm, $d_{sep}$ =400 nm and outer clad diameter of 10 μm at SF=1.7, Inset: Von Mises stress distribution in the respective structures.

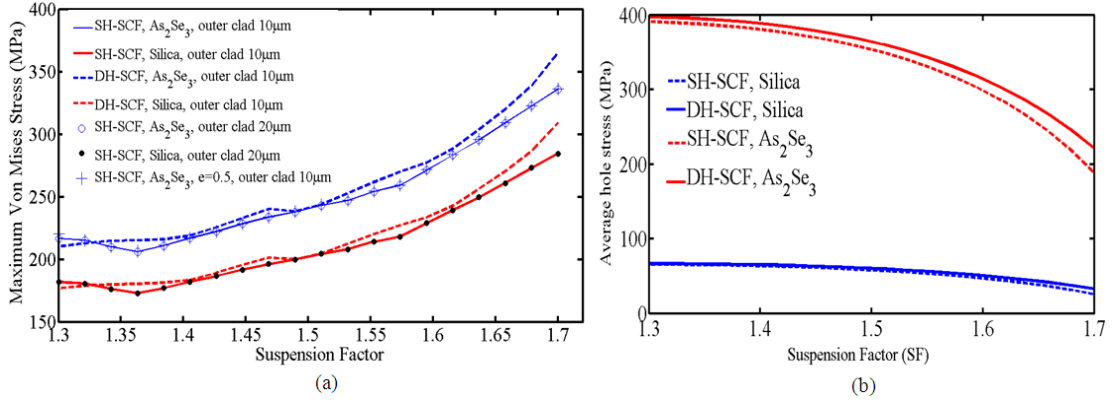

Fig. 3. (a) Effect of suspension factor (SF) on maximum Von Mises Stress for two different outer clad diameters. $e$ is the parameter controlling ellipticity of the air hole. $e=1$ for purely circular hole. (b) Average single hole stress for an outer clad diameter of 10 μm and SF=1.7, $d_{bore}$=100 nm.

### 4.2 Properties of fundamental mode cutoff (FMC)

The $HE_{11}$ mode profiles as shown in Fig. 4(a), (b), (c) confirm that, guided mode can be obtained in such structures with non-gaussian distribution. Here, the field enhancement at the interface of central air holes and fiber material is governed by the ratio $\varepsilon_{fiber}/\varepsilon_{Air}$. For $As_2Se_3$, this ratio can be as high as 9. To confirm the additional enhancement provided by the suspended air holes in the surrounding, we have investigated the field patterns when the nano hole of the same size is inside an $As_2Se_3$ nanowire (blue line in Fig. 4(d)). The peak electric field can be enhanced by about 95% by placing air hole in the core of the SCF instead of placing it inside an $As_2Se_3$ wire. Such impact of suspended holes will not change much when the fiber material is silica instead of $As_2Se_3$. Hence, in this section, we have limited our study to $As_2Se_3$ only. Although the peak intensity of electric field in DH structure is equal to that in SH structure, it decays inside the holes of DH-SCF at a faster rate. However, such trapping of light inside the DH structure can be improved by optimizing the bore diameters carefully. In fact, the fundamental propagation mode of the proposed structures is highly sensitive to bore diameter and the operating wavelength as described in the following subsections.

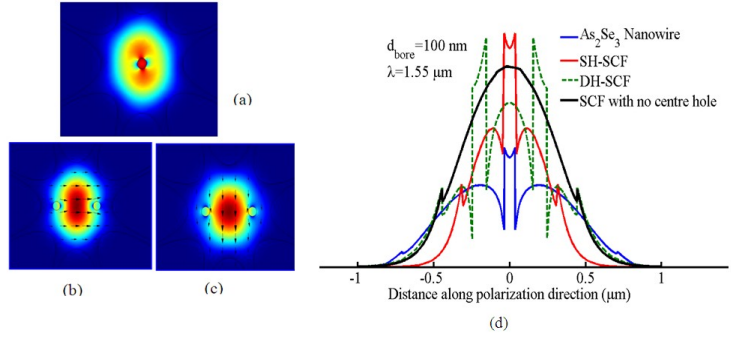

Fig. 4. $HE_{11}$ mode profile in (a) SH-SCF (b) DH-SCF, x-polarization (c) DH-SCF, y-polarization at SF=1.7. The operating wavelength is 1.55 µm. Trapping of light is clearly visible for the SH-SCF structure. Here $d_{bore}$=100 nm and $d_{sep}$ is 400 nm (d) Comparative improvement of field enhancement inside the nanoholes for the proposed structures at SF=1.7.

### 4.2.1 Wavelength dependence

In Fig. 5(a), the mode effective area versus wavelength has been shown as a function of SFs in SH and DH- structures. The corresponding hole power fractions are also illustrated in Fig. 5(b). We have found that, for a fixed value of SF, $d_{bore}$ and $d_{sep}$, there exists a lower cutoff wavelength at which the fundamental mode disappears. We have investigated such unusual condition at different SFs and different bore diameters to confirm that, this is not due to any simulation artifact. A definite relation of cutoff condition with suspension factor and bore diameter has been explored observing the simulation results. However, a rigorous modeling of such phenomenon is beyond the scope of this work. Such fundamental mode cutoff condition (FMC) has been reported before in [25] where a silica core PCF having $GeO_2$ doped cladding has been proposed. The large effective area PCF has been designed by exploiting the effect of increasing cladding index ($n_{cl}$/ $n_{FSM}$) beyond the core index precisely. On the contrary, the proposed designs of this work have significantly smaller effective areas suitable for low power nonlinear applications. It is important to be mention here that, presence of air holes at the core region effectively lessens the core index of the suspended core fiber. At the cutoff condition, we have observed a sharp transition of effective area and effective index which is an indication of mode disappearance. Effective area method has been taken into account to characterize upper cutoff wavelength already in [24]. In addition, the power confined inside the central holes/bores falls to zero drastically beyond the cutoff wavelength which basically implies absence of any fundamental guided mode in the structures. Confinement loss of several orders of magnitudes can be monitored at such condition by utilizing a properly designed perfectly matched layer. It has been found that lower cutoff wavelength in DH structure is larger than that in SH for a particular SF. Again, increasing the SF will tend to shift the cutoff wavelength towards higher value for a particular structure. In SH-SCF, cutoff wavelength is 900 nm and 1300 nm for SF=1.65 and SF=1.7, respectively. In Fig. 5(b), it can be seen that, hole power fraction increases with the increase in wavelength and at cutoff, it decreases to zero drastically. Besides, with the increase in SF, the power fraction in both the structures also increases. Furthermore, two distinct FMC wavelengths exist for the birefringent DH-structure at a certain SF. In sum cut-off condition shifts to longer wavelength as the distance between the central hole and deformed air hole decreases in both the structures. Such phenomenon of exhibiting a lower cutoff has not been observed for the proposed SCF having no holes in the core region [26]. This effect may be studied further by exploiting the principles governing the operation of slot waveguide. In a slot guide, slot dimension should be effectively less than the decay length of evanescent field for obtaining a guided mode in the slot region. In our design, the suspended air holes and the central holes are jointly forming multiple slot like regions where field enhancement occurs due to large index contrast. We have found that, by changing SF or $d_{bore}$ at a particular $d_{sep}$, the cutoff condition can be tuned which essentially implies the existence of certain characteristic decay length here. It can be concluded here that, both the fundamental core mode and power fraction inside the bores disappear simultaneously and so substantiates the role of such air holes on achieving fundamental mode cutoff condition in suspended core fibers.

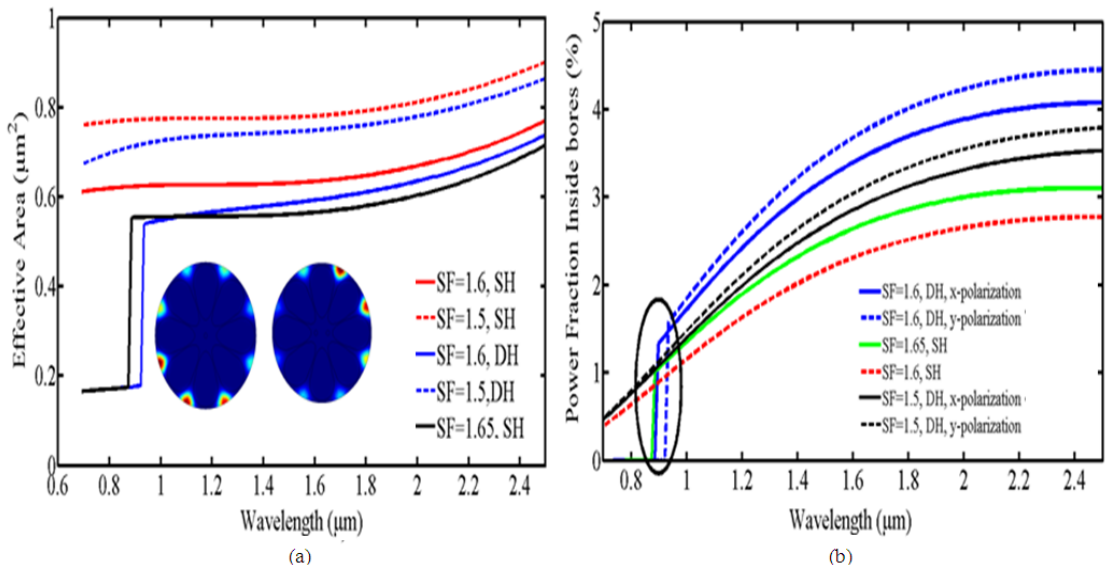

Fig. 5. (a) Effect of suspension factors and structures on lower cutoff wavelength. Here $d_{bore}$ =100 nm. The sharp transition of effective area denotes the fundamental mode cutoff condition in the proposed structures. Inset: Absence of guided mode below the cutoff wavelength. (b) variation of effective hole power fraction at $HE_{11}$ mode.

### 4.2.2 Effects of nanoholes

The proposed geometries exhibit high degree of sensitivity to bore diameter at certain SF too. There exists a certain fundamental mode cutoff diameter at a particular wavelength when SF is increased beyond a certain limit. In Fig. 6, variation of power fraction inside the nanoholes are shown at 1.55 µm. The rapid fall of hole power fraction indicates the existence of cutoff diameter. This is again confirmed by the sharp transition of effective area as shown in Fig.7. At this condition, the effective index becomes nearly constant and the structures possess no guided mode. However, the cutoff diameter in DH-SCF structure is significantly smaller than that in SH-SCF structure. Besides, the polarization dependence of cutoff diameter for the birefringent DH-SCF structure is shown in Fig. 6(b). It should be noted that, at a certain value of SF there exists a certain bore diameter in DH-SCF structure, for which it behaves like a single polarization device allowing only one mode of polarization to exist. In summary, at larger SF, cutoff diameter shifts to smaller value in both the structures. This is due to the fact that, at larger SF the distance between central air holes and suspended air holes gets much smaller than the characteristic decay length.

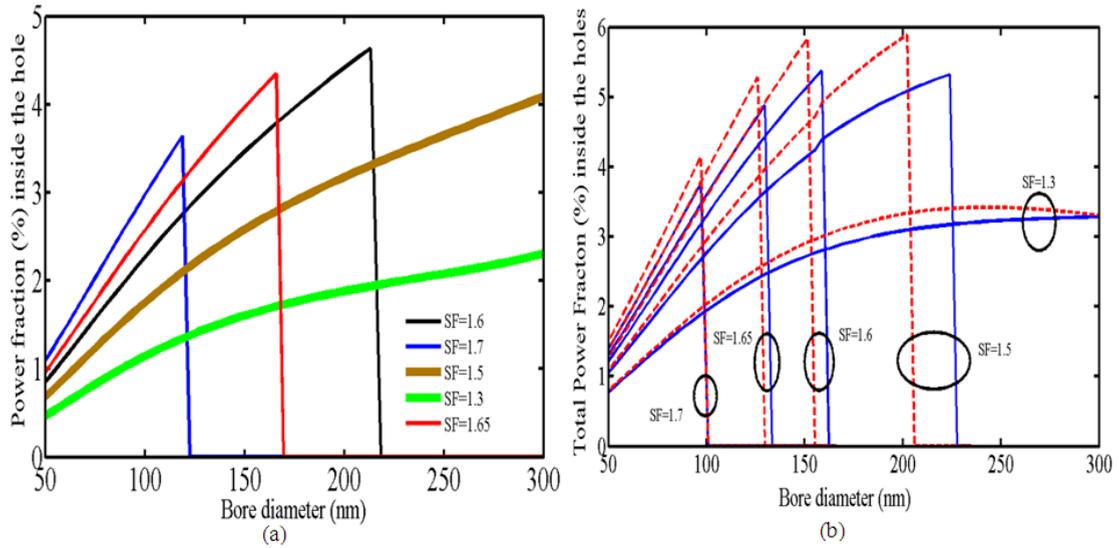

Fig. 6. Shifting of Fundamental mode cutoff bore diameter for different suspension factors at an operating wavelength of 1.55 μm in (a) SH-SCF (b) DH-SCF (Dotted and solid lines for two orthogonal polarizations of the fundamental mode.) Maximum core diameter is 1.35 μm at SF=1.3.

**4.3 Properties of sensing**

The SCFs has been adopted extensively as highly sensitive evanescent biosensor. It has been reported that, more than 80% of modal power can exist inside the deformed holes filled with water for a core diameter of 200 nm [27]. Here, we have investigated the sensitivity coefficient defined in [28] for the proposed geometries considering water as the sample material ($n_r$=1.33). Results are shown in Fig. 8(a) and in Fig. 8(b). Here, we have only considered $As_2Se_3$ to study the sensing property. It is to be noted that, no significant improvement of %r as defined by equation 7 has been observed in DH-structure over SH-structure in our analysis. However, %r in hole assisted SCF has been found to be larger than that in normal SCF based sensor for certain values of SF. It has been found that, power confined in the sample media increases due to the presence of hole since the rate of leakage from the core increases in this case. The modal effective index decreases as the light interacts more with air during propagation and the sensitivity coefficient improves eventually. Again, with the increase in hole diameter %r increases as shown in Fig. 8(b) due to the reduction of effective index and enhancement of power leakage from the core region. The flat portion of the curve indicates unguided mode in the SCF sensor in presence of water in the deformed holes. Besides, the sensitivity coefficient improves with the increase in SF as the interaction between the core field and sample material becomes stronger. Thus, it can be mentioned that, the proposed SFs and the air hole region within the core may be treated as tuning parameters for SCF based sensor devices. However, comparison of such sensitivity parameter of the proposed structure having guided region of comparatively larger dimension with other platforms including slot waveguides, nanowires and small core micro-structured fibers is beyond this work.

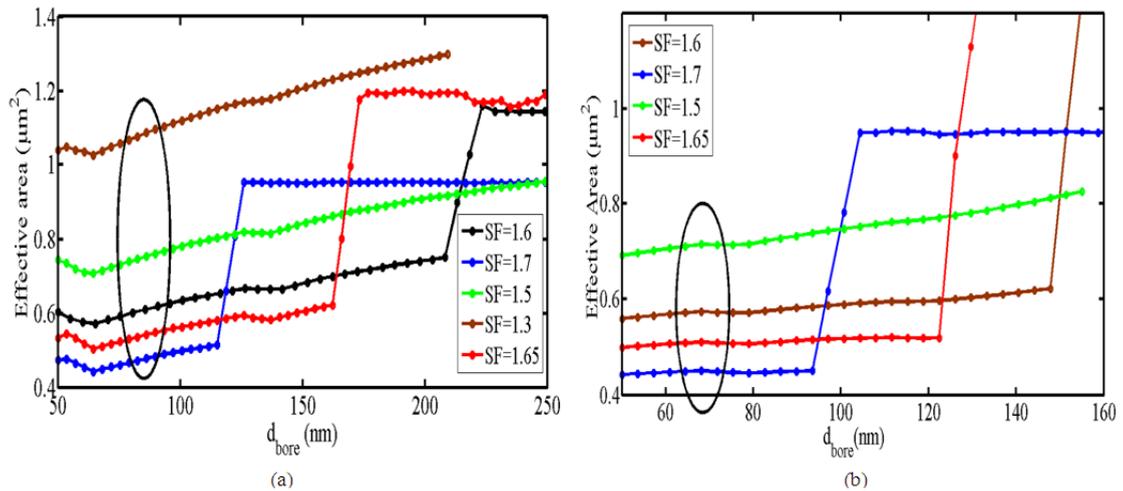

Fig. 7. Variation of effective area with suspension factor at a wavelength of 1.55 μm in (a) SH-SCF (b) DH-SCF.

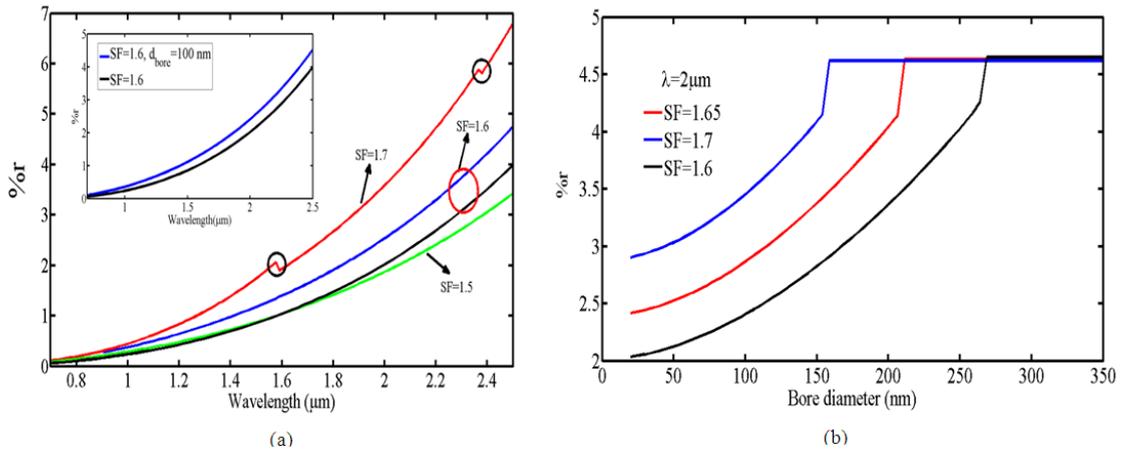

Fig. 8. (a) Sensitivity variation with wavelength. Black line corresponds to SCF with no air hole. The discontinuity in red line corresponds to FMC at SF=1.7 in DH structure. Inset: %r variation in SH-SCF and SCF structure. (b) Variation of %r with $d_{bore}$ in SH-structure. Flat region corresponds to FMC condition.

### 4.4 Group Velocity Dispersion, Mode mismatch and Birefringence

In Fig. 9(a), group velocity dispersion (GVD) profiles for $As_2Se_3$ background have been shown. The dispersion profiles in SH-SCF and DH-SCF structure are almost identical here. It is to be noted that, although the confinement of light in core is around 90%, the maximum confinement in hole is about 5-6% for both the structures. Hence, the GVD curves look similar for both the cases. Here, the first zero dispersion wavelength shifts towards the lower wavelengths with the increment in SF. Two zero dispersion wavelengths can also be achieved in the same structures if the core diameter is reduced significantly [26]. The % overlap shown in Fig. 9(b) shows that, the proposed structures have low coupling efficiency at a particular wavelength for conventional SMF coupling. It improves as the confinement reduces i.e mode field radius increases through the reduction of SF or by considering a fiber material of lower index. Coupling efficiency can be improved by introducing high performance grating or inverted fiber coupler [29]. Later, phase birefringence ($B_p$) of the proposed geometries ignoring any photo induced or thermally induced anisotropy of $As_2Se_3$ has been investigated. Refined meshing has been performed to avoid any artifact induced birefringence. It has been found that, the SH structure has almost zero birefringence. However, birefringence can be increased significantly by distorting the circular shape of the centre hole as shown in Fig. 9(c). Here, ellipticity has been introduced by shrinking the x-axis of the central hole. Such ellipticity has been found to affect the FMC condition at a particular SF too. However, DH structure is inherently birefringent as shown in Fig. 9(d). Suspension factor can influence $B_p$ in DH structure also. Highly birefringent evanescent sensors have been proposed in [30], [31] very recently.

## 5. Conclusion

In this work, a novel design of single mode suspended core fiber possessing short wavelength fundamental cutoff property has been proposed. Such waveguide structure is ideally useful for designing optical long pass filters suitable for different applications. Plane strain analysis of the proposed structures has numerically compared the limitations imposed by thermal stress likely to be generated during the fabrication on different platforms. The cutoff property has been thoroughly investigated and a definite relation between the geometrical parameters and cutoff wavelength has been explored. In addition, improvement of fiber birefringence and evanescence sensing efficiency has been numerically demonstrated due to the presence of such air holes inside the SCF. In sum, such theoretical investigation will lay a solid ground for future study of mechanical and optical properties sub

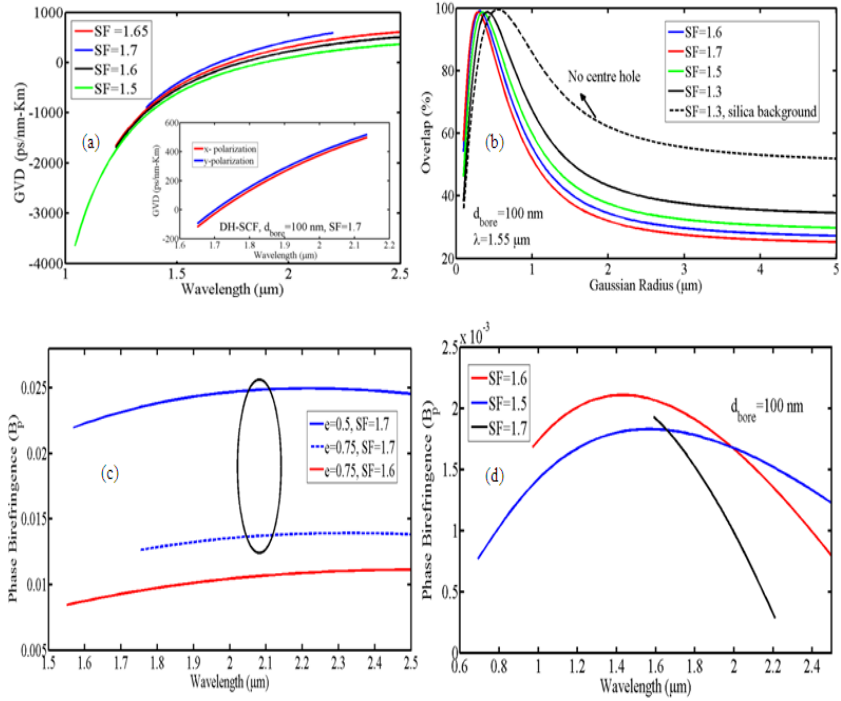

Fig.9 (a) GVD in SH structure at $d_{bore}$=100 nm. Inset: GVD in DH structure for two fundamental mode polarizations. (b) %overlap integral in As2Se3 and $SiO_2$- SH structure (c) Variation of birefringence in $As_2Se_3$ SH structure at $d_{bore}$=100 nm for elliptical central hole (d) Birefringence in $As_2Se_3$ DH structure at wavelengths greater than the FMC wavelength.

wavelength core suspended core fiber in presence of additional nano scale air regions which have potential engineering merits in wide range of applications.

## References

[1] V. S. Afshar, W.Q.Zhang, H. Ebendorff-Heidepriem, and T. M. Monro, "Small core optical waveguides are more nonlinear than expected: experimental confirmation," *Opt. Lett.*, vol. 34, no. 22, pp. 3577-3579, Nov. 2009.

[2] G. S. Wiederhecker, C. M. b. Cordeiro, F. Couny, F. Benabid, S. A. Maier, J. C. Knight, C. H. b. Cruz, and H. L. Fragnito, "Sub-Wavelength Intensity Profiles and Field Enhancement within an Optical Fiber," in Proceedings of conference on Lasers and Electro-Optics/Quantum Electronics and Laser Science Conference and Photonic Applications Systems Technologies, 2007.

[3] Y. Ruan, H. Ebendorff-Heidepriem, S. Afshar, and T. M. Monro, "Light confinement within nanoholes in nanostructured optical fibers,"*Opt. Expr*., vol. 18, no. 25, pp. 26018-26026, 2010.

[4] A. Agrawal, N. Kejalakshmy, B. M. A. Rahman, and K. T. V. Grattan, "Soft Glass Equiangular Spiral Photonic Crystal Fiber for Supercontinuum Generation," *Photon. Technol. Lett.*, vol. 21, no. 22, pp. 1722 -1724, Nov. 2009.

[5] M. N. Hossain, M. S. Alam, D. M. N. Hasan, and K. M. Mohsin, "A Highly Nonlinear Spiral Photonic Crystal Fiber for Tailoring Two Zero Dispersion Wavelengths in the Visible Region," *Photon. Lett. Poland*, vol. 2, no. 3, pp. 143-145, 2010.

[6] G. S. Wiederhecker, C. M. B. Cordeiro, F. Couny, F. Benabid, S. A. Maier, J. C. Knight, C. H. B. Cruz, and H. L. Fragnito, "Field enhancement within an optical fiber with a subwavelength air core," *Nature Photonics,* vol. 1, pp. 115-118, Feb. 2007.

[7] V. R. Almeida, Q. Xu, C. A. Barrios, and M. Lipson, "Guiding and confining light in void nanostructure," *Opt. Lett.*, vol.29, no. 11, pp. 1209-1211, June 2004.

[8] I. Alam and K. Hamamoto, "Slot Waveguide by Using Double High-Mesa Structure for Optical Absorption Sensing," *Jpn. J. Appl. Phys.*, vol.49, Dec. 2010.

[9] L. Vivien, D. Marris-Morini, A. Griol, K. B. Gylfason, D. Hill, J. Álvarez, H. Sohlström, J. Hurtado, D. Bouville, and E. Cassan, "Vertical multiple-slot waveguide ring resonators in silicon nitride", *Opt. Exp.*, vol. 16, no. 22, Oct. 2008.

[10] J. K. Chandalia, B. J. Eggleton, R. S. Windeler, S. G. Kosinski, X. Liu, and C. Xu, "Adiabatic Coupling in Tapered Air-Silica Microstructured Optical Fiber," *Photon. Technol. Lett.*, vol. 13, pp. 52-54, May 2001.

[11] G. E. Farin, and H.Gerald, *Curves and surfaces for computer-aided geometric design*, 4th ed., Elsevier Science & Technology Books, 1997.

[12] A. S. Webb, F. Poletti, D. J. Richardson, and J. K. Sahu, "Suspended-core holey fiber for evanescent-field sensing," *Opt. Eng.*, vol. 46, no. 010503 ,Jan. 26, 2007.

[13] T. Monro, S. Afshar, H. Ebendorff-Heidepriem, W. Q. Zhang, and Y. Ruan, "Emerging Optical Fibers: New Fiber Materials and Structures," Conference on Lasers and Electro-Optics (CLEO), Baltimore, Maryland, May 31, 2009.

[14] Q. Coulombier, L. Brilland, P. Houizot, T. Chartier, T. N. N'Guyen, F. Smektala, G. Renversez, A. Monteville, D. Méchin, T. Pain, H. Orain, J. Sangleboeuf, and J. Trolès, "Casting method for producing low-loss chalcogenide microstructured optical fibers," *Opt. Exp.,* vol. 18, pp. 9107-9112, April 2010.

[15] H. Ebendorff-Heidepriem, and T. Monro, "Extrusion of complex preforms for microstructured optical fibers," *Opt. Exp.*, vol. 15, pp. 15086-15092, Oct. 2007.

[16] M. Liao, C. Chaudhari, X. Yan, G. Qin, C. Kito,T. Suzuki, and Y. Ohishi, "A suspended core nanofiber with unprecedented large diameter ratio of holey region to core", *Opt. Exp.*, vol. 18, no. 9, pp. 9088-9097, 2010.

[17] COMSOL Multiphysics, Version 3.2, 2005.

[18] M. J. Weber, *Handbook of Optical Materials*, Lawrence Berkeley National Laboratory, University of California, Berkeley, California, 2003.

[19] B. Ung and M. Skorobogatiy, "Chalcogenide microporous fibers for linear and nonlinear applications in the mid-infrared," *Opt. Exp*. vol.18, pp. 8647-8659, April 2010.

[20] H. C. Nguyen, B. T. Kuhlmey, E. C. Mägi, M. J. Steel, P. Domachuk, C. L. Smith, and B. J. Eggleton, "Tapered photonic crystal fibres: properties,characterisation and applications," *Appl. Phys. B*, vol. 81, pp.377–387, Oct. 2005.

[21] M. A. Foster and A. L. Gaeta, "Ultra-low threshold supercontinuum generation in sub-wavelength waveguides," in Proceedings of Frontiers of Optics, Rochester, New York, Oct., 2004.

[22] D. J. McEnroe and W. C. LaCourse, "Tensile Strengths of Se, $As_2S_3$, $As_2Se_3$, and $Ge_{30}As_{15}Se_{55}$ Glass Fibers," *Journal of the American Ceramic Society*, vol.72, no.8, pp. 1491–1494, Aug. 1989.

[23] T. Schreiber, H. Schultz, O. Schmidt, F. Röser, J. Limpert, and A. Tünnermann, "Stress-induced birefringence in large-mode-area micro-structured optical fibers," *Opt. Exp.*, vol. 13, no. 10, pp. 3637-3646, May 2005.

[24] J. Chianga and T. Wu, "Analysis of propagation characteristics for an octagonal photonic crystal fiber (O-PCF)," *Opt. Comm.*, vol. 258, no. 2, pp. 170-176, 15 Feb. 2006.

[25] C. J. S. de Matos, "Modeling Long-Pass Filters Based on Fundamental-Mode Cutoff in Photonic Crystal Fibers," *Photon. Technol. Lett.*, vol. 21, no. 2, Jan. 15 2009.

[26] K. M. Mohsin, M. Shah Alam, D. M. N. Hasan, and M. N. Hossain, "Dispersion and nonlinearity properties of a chalcogenide $As_2Se_3$ suspended core fiber," *Appl. Opt.*, vol. 50, no. 25, pp. E102-E107, Sept., 2011.

[27] T. M. Monro, S. Warren-Smith, E. P. Schartner, A. François, S. Heng, H. Ebendorff-Heidepriem, S.V. Afshar, "Sensing with suspended-core optical fibers," *Opt. Fiber Technol.*, vol. 16, pp.343–356, Oct. 2010.

[28] C. M. B. Cordeiro, M. A. R. Franco, G. Chesini, E. C. S. Barretto, R. Lwin, C. H. Brito Cruz, and M. C. J. Large, "Microstructured-core optical fiber for evanescent sensing applications," *Opt. Exp.*, vol. 14, no.26, pp.13056–13066, Dec., 2006.

[29] J. V. Galan, P. Sanchis, J. Blasco, and J. Marti, "Horizontal slot waveguide-based efficient fiber couplers suitable for silicon photonics," in Proceedings of ECIO conference, Eindhoven, 2008.

[30] J. Kou, F. Xu, and Y. Lu, "Highly Birefringent Slot-microfiber," *Photon. Technol. Lett.*, vol. 23, no.15, pp. 1034 - 1036, July 2011.

[31] Y. Jung, G. Brambilla, K. Oh, and D. J. Richardson, "Highly birefringent silica microfiber," *Opt. Lett.*, vol. 35, no. 3, pp. 378-380, Jan. 2010.